\documentclass{nature}
\usepackage{graphicx}
\usepackage{amssymb}
\usepackage{dcolumn}
\usepackage{bm}
\usepackage{epstopdf}
\usepackage{epsfig}
\usepackage{color}
\usepackage[colorlinks=true, letterpaper=true, pdfstartview=FitV, linkcolor=blue, citecolor=blue, urlcolor=blue]{hyperref}

\begin{document}

\title{A new kind of 2D topological insulators BiCN with a giant gap and its substrate effects}

\author{Botao Fu$^{1}$, Yanfeng Ge$^{1}$, Wenyong Su$^{1}$, Wei Guo$^{1}$,Cheng-Cheng Liu$^{1,\star}$}

\maketitle

\begin{affiliations}
\item School of Physics, Beijing Institute of Technology, Beijing 100081, China

$^\star$e-mail: ccliu@bit.edu.cn
\end{affiliations}

\begin{abstract}
Based on DFT calculation, we predict that BiCN, i.e., bilayer Bi films passivated with -CN group, is a novel 2D Bi-based material with highly thermodynamic stability, and demonstrate that it is also a new kind of 2D TI with a giant SOC gap ($\sim 1$ eV) by direct calculation of the topological invariant $Z_2$ and obvious exhibition of the helical edge states. Monolayer h-BN and MoS$_2$ are identified as good candidate substrates for supporting the nontrivial topological insulating phase of the 2D TI films, since the two substrates can stabilize and weakly interact with BiCN via van der Waals interaction and thus hardly affect the electronic properties, especially the band topology. The topological properties are robust against the strain and electric field. This may provide a promising platform for realization of novel topological phases.
\end{abstract}

Two-dimensional (2D) topological insulators (TIs), aka quantum spin Hall (QSH) insulators, have attracted numerous interest in material science and condensed matter physics due to its scientific importance as an unique  symmetry protected topological (SPT) quantum state and its potential technological applications ranging from spintronics to topological quantum computation~\cite{Hasan2010,Qi2011}.
This novel electronic state has a bulk gap but can conducts charge and spin current without dissipation via the spin-momentum locked gapless edge state protected by time-reversal symmetry. The prototypical 2D TI was first proposed in graphene~\cite{kane2005a,kane2005b}, in which the spin-orbit coupling (SOC) opens a band gap at the Dirac points. However, the rather tiny second-order effective SOC makes the 2D TI state in graphene only appear at an unrealistically low temperature~\cite{yao2007,Min2006}. So far, the QSH effect is only experimentally verified in HgTe/CdTe~\cite{Bernevig2006,Science.318.766} and InAs/GaSb~\cite{Liu2008,PRL.Knez.2012} quantum wells, in stringent conditions, e.g., ultrahigh-quality samples and ultralow temperature, due to their small bulk band gaps (at the order of meV).
Therefore new 2D TIs with large bulk gaps  which can realize QSH effect easily  are still much desired.

Extensive effort has been devoted to the search for new QSH insulators with a large SOC gap~\cite{Murakami2006,liu_quantum_2011,liu_low-energy_2011,Liu2011,Hirahara2011,Xu2013,Yang2012,Song2014,Liu2014,Zhou2014,Wang2014,Ma2015,Zhang2012,Qian2014,Huang2013,Weng2014,Zhou2015}.
For instance, honeycomb lattice type materials such as silicene, germanene~\cite{liu_quantum_2011} or stanene~\cite{liu_low-energy_2011} , and chemically modified stanene~\cite{Xu2013} have been proposed. The element bismuth (Bi) has the largest SOC strength in the periodic table except radioactive elements. Therefore, the above exotic QSH effect can be expected to emerge notably in Bi-based materials. The bilayer Bi film has long been predicted as TI, with an inverted SOC gap of 0.2 eV at $\Gamma$ point~\cite{Murakami2006,Wada2011,Koroteev2008}, which can be described by BHZ model~\cite{Bernevig2006}. In addition, chemical modification provides excellent control means to improve the key properties of systems with the relevant physics altered at the same time. For example, when the bilayer Bi film are hydrogenated or halogenated from both sides, the stable 2D honeycomb Bi hydride (Bismuthumane) and halide can be obtained, which can be described by modified Kane-Mele model~\cite{kane2005a,Liu2014}. The SOC gap at K and K' can reach the recorded 1 eV~\cite{Song2014,Liu2014}, much larger than that of bilayer Bi films~\cite{Wada2011,Koroteev2008}. Besides, chemical group Methyl (-CH3) is used to modified the bilayer Bi film~\cite{Ma2015}. Here we propose cyano (-CN) as another chemical group to tune the bilayer Bi film by passivating every Bi atom with a -CN group from both sides or one side. We find the two BiCN monolayers (regardless of the passivation from both sides or one side) are 2D TIs with a huge SOC gap of approximate 1 eV. The low-energy effective Hamiltonian is developed for the symmetric one (passivation from both sides). Moreover, we investigate monolayer h-BN and MoS$_2$ as candidate substrates, and find the composite systems are van der Waals (vdW) heterostructures.

\section*{Results}
\subsection{Structure and stability.}
\begin{figure}
  \begin{center}
  \epsfig{file=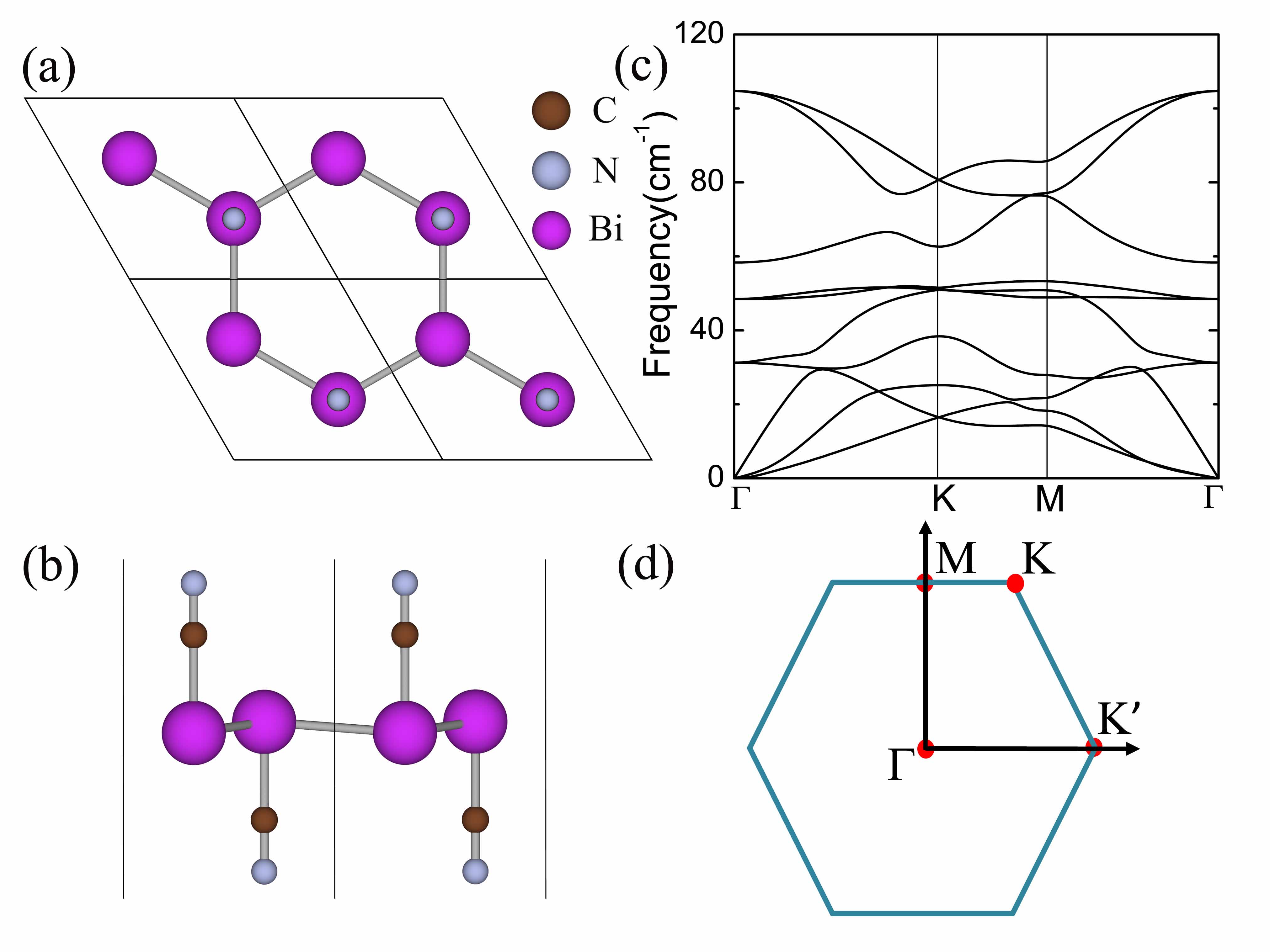,width=\textwidth}
\caption{Geometric structure and Phonon spectrum. (a),(b) Optimized geometric structure of BiCN from top and side views. (c) Phonon spectrum of BiCN. (d) The FBZ (First Brillouin zone) and related high-symmetry k points.\label{fig:geometry}}
\end{center}
\end{figure}

The geometric structure of BiCN is shown in Fig.~\ref{fig:geometry}(a),(b). This new 2D material can be regarded as a freestanding high-buckled hexagonal bilayer Bi film functionalized by the cyano groups(-CN) from top and bottom sides. This functionalization will greatly modify the buckling structure and electronic properties as well. The optimized lattice constant of BiCN is 5.54 $\AA$ and the buckling height of Bi atoms is only 0.25 $\AA$. The in-plane bond length of two nearest Bi atoms is 3.21 $\AA$ and the out-plane bond length of Bi and -CN is 2.23 $\AA$. The computed phonon spectrum is shown in Fig.~\ref{fig:geometry}(c). There are eighteen phonon modes for six atoms in the unit cell, but for a better view we here just show ten phonon modes with the lowest frequency. The absence of imaginary vibrational frequency over the whole First Brillouin Zone (FBZ) confirms the dynamic stability of this system.

\begin{figure}
  \begin{center}
  \epsfig{file=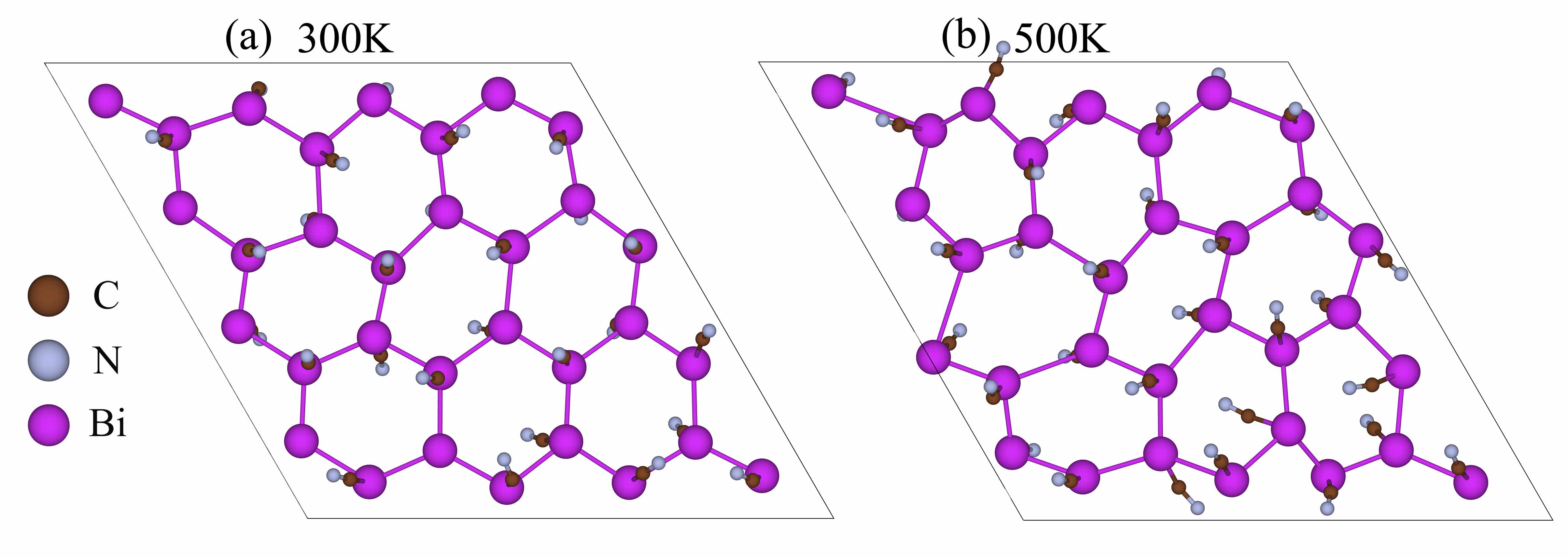,width=\textwidth}
\caption{Snapshot taken from MD simulation for 4$\times$4 supercell of BiCN monolayer at temperature of 300K(a) and 500K(b) after 2.3 ps.\label{fig:md}}
\end{center}
\end{figure}

The thermodynamic stability is also confirmed by first principle molecular dynamics (MD) simulations and the evaluation of formation energy. We use a 4$\times$4  supercell to perform the MD simulation at 300K and 500K respectively. Random chosen samples of geometric structure after MD running for 2.3 ps are given in Fig.~\ref{fig:md}. We can see the honeycomb of Bi atom is well maintained at 300K while the -CN functional groups have a little swing due to the thermal perturbation. Moreover, even at 500K, the structure of BiCN is still unbroken.
The formation energy of BiCN is defined as
${{E}_{f}}={{E}_{BiCN}}-{{n}_{Bi}}{{E}_{Bi}}-{{n}_{C}}{{E}_{C}}-{{n}_{N}}{{E}_{N}}$
, where $E_{BiCN}$ is the total energy of BiCN monolayer and $E_{Bi},E_{C},E_{N}$ are the chemical potential of Bi, C, N atoms using bulk bismuth, diamond, nitrogen molecules, respectively. The  $n_{Bi},n_{C},n_{N}$  are the numbers of Bi, C, N atoms in the unitcell. The calculated formation energy of BiCN is about -8.18 eV per unitcell, which is much larger then BiCH$_{3}$(-0.52 eV per unitcell)~\cite{Ma2015}, BiH(0.92 eV per unitcell)~\cite{Song2014}, BiF(-4.92 eV per unitcell)~\cite{Song2014}, BiCl(-2.04 eV per unitcell)~\cite{Song2014}, and BiI(-1.04  eV per unitcell). These indicate BiCN has the most highly thermodynamic stability among the functionalized Bismuth bilayers.


\subsection{Electronic structure, topological properties and helical edge states.}
\begin{figure}
  \begin{center}
  \epsfig{file=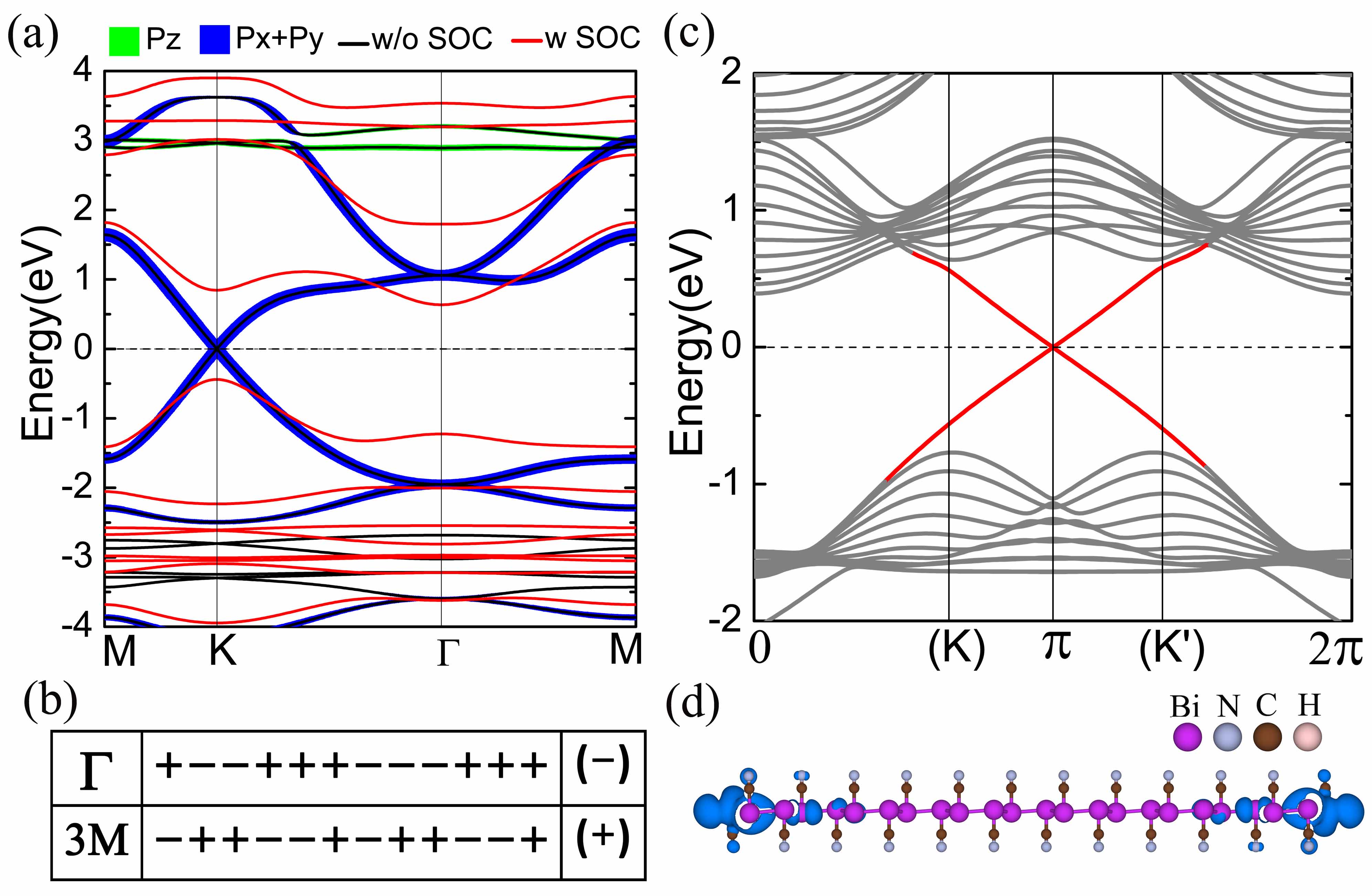,width=\textwidth}
\caption{Electronic structure and topology. (a) The bandstructure of intrinsic BiCN with SOC (red line) and without SOC (black line). The green and blue line indicates the $p_z$ and $p_x/p_y$ orbital character respectively. The thicker line means the greater weight of certain orbital. (b) The parity table of the twelve pairs of the occupied double-degenerated valence bands at $\Gamma$ and three M points. The positive and negative values stand for even and odd parity respectively. The product of those parities at each k is given at the right brackets. (c) Bandstructure of zigzag nanoribbon with width of N=12 (5.44 nm). The red line represents the helical edge state and the gray line represents the bulk state. The K (K')  in brackets means the 1D projection of K(K') in 2D BZ. (d) Charge density distribution in real space of the edge state near $\pi$ point shown in (c).\label{fig:bs}}
\end{center}
\end{figure}

Figure~\ref{fig:bs}(a) shows the bandstructures of BiCN with SOC (red color) and without SOC (black color). In the absence of SOC, the system is a graphene-like Dirac semimetal with valence and conduction bands touching each other at the K (K') points in the FBZ, forming Dirac cone with isotropic linearly dispersion relation. The Fermi level just crosses the Dirac points and make it a typical Dirac semimetal with the Fermi velocity $v_f=7.6*10^5$ m/s. The atomic orbital projections of the band near the Dirac points show that the main band components come from the $p_x/p_y$ orbitals of Bi atoms. It is greatly different from that of graphene whose Dirac cone is formed by the half-occupied $p_z$ orbital in a honeycomb structure. When SOC is turned on, the original gapless Dirac cone is opened and the system becomes a TI with a huge non-trivial gap of 1.28 eV at K point and indirect gap of 1.04 eV between conduction band maximum (CBM) at $\Gamma$ point and valence band maximum (VBM) at K point. Due to the existence of space inversion symmetry for BiCN system, we can simply calculate the topological invariant by the parity method proposed by Fu and Kane~\cite{Fu2007}.
The topological invariant index $\nu=0$ indicates a topological trivial phase while $\nu=1$ means a topological nontrivial phase. Here $\nu$ is define as
\begin{equation}\label{z2}
\left(-1\right)^{\nu}=\prod_{i=1}^{4}\delta\left(k_{i}\right)=\delta\left(\Gamma\right)\delta\left(M\right)^{3},
\end{equation}
where $\delta\left(k_{i}\right)=\prod_{n\in val}^{N}\xi_{2n}^{i}$,  $\xi_{2n}^{i}$ is the parity of the 2n$^{th}$ occupied eigenstate at time-reversal invariant point (TRIP) $k_i$, and N is the number of the Kramers pairs of  valence bands. The parity table of twelve pairs of the occupied valence bands at $\Gamma$ and three M points is shown in Fig.~\ref{fig:bs}(b). From total parities we obtain topological index $\nu=1$, which explicitly proves BiCN to be a 2D TI.

For structures with a finite boundary, 2D TIs have odd number pairs of helical edge states traversing the bulk band gap. These kinds of edge states can provide dissipationless conducting channels at boundary of a sample and be protected by the topological invariance. To further demonstrate the topological properties of BiCN system we calculate the electronic structure of its zigzag nanoribbon with width of 5.44 nm (N=12). To eliminate the effect of dangling bonds on the edge, the marginal Bi atoms are saturated by hydrogen atoms. Figure~\ref{fig:bs}(c) displays the bandstructure of the zigzag nanoribbon. We can clearly see one pair of edge state (red lines) traverse the bulk gap, cross each other and form a Dirac point at the middle of 1D BZ. Each state of the pair transports in opposite direction with filtered spin due to the protection of time reverse symmetry. Besides, the ribbon has two inversed-symmetric edges, so each edge state above has its degenerate copies for another edge. Figure~\ref{fig:bs}(d) depicts the charge density distribution of edge states near the Dirac point, which clearly shows they are well-localized along the edge of the ribbon.

In the following, we build a four-band model to capture the low-energy physics of BiCN.
Akin to the bilayer Bi film with hydrogenated or halogenated from both sides~\cite{Liu2014}, the symmetry-adopted basis functions can be written as linear combination of $p_x$ and $p_y$ orbitals of Bi atoms, i.e. $-\frac{1}{\sqrt{2}}\left(p_{x}^{A}+i\tau_{z}p_{y}^{A}\right) $, $\frac{1}{\sqrt{2}}\left(p_{x}^{B}-i\tau_{z}p_{y}^{B}\right)$ in the symmetric BiCN, with two distinct sites A and B in a unit cell, and $\tau_{z}$ as valley index (K or K'). Expanding the Hamiltonian to the first order of k around K (K') point, the symmetry-allowed four-bands low-energy effective model involving SOC can be written as
\begin{equation}\label{H}
H_{\tau}=v_{F}\left(k_{x}\sigma_{x}+\tau_{z}k_{y}\sigma_{y}\right)+\lambda_{so}\tau_{z}\sigma_{z}s_{z},
\end{equation}
where Pauli matrix $\sigma$, $\tau$ and $s$ denote A, B sublattices, valley index, and spin respectively. The second term is the on-site SOC, which exactly plays a crucial role to develop a topological nontrivial phase with a giant gap. The above low-energy effective Hamiltonian is invariable under the space inversion operation and the time reversal operation. The only two parameters $v_f=7.6*10^5$ m/s and $\lambda_{so}=0.64$ eV are determined by fitting with the DFT calculations.

So far we have demonstrated BiCN to be a new kind of 2D TIs, and clarified the related physics. There are some advantages need to be emphasized here. Firstly, BiCN is saturated by cyano groups(-CN) on both sides, which may protect it from being destroyed by environment. Secondly, the giant non-trivial band gap is induced by the on-site SOC of Bi, very robust against the external strain and electric field as well as finite temperature (details at part II and I of Supplemental Material). Thirdly, the large gap of 1.04 eV can effectively suppress the emergence of heat-activated carriers from bulk state and ensure the realization of QSH effects at room temperature. Besides, the giant gap will facilitate the localization of edge states and avoid the coupling of the edge states from different edges for narrow nanoribbons~\cite{Zhou2008} . Last, with graphene-like hexagonal structure, the electronic structure of BiCN is valley related, which may have broad applications in valleytronics.

\subsection{Substrate effects.}
Recently several 2D TIs are predicted, such as low-buckled silicene~\cite{liu_quantum_2011}, stanene~\cite{liu_low-energy_2011}, bilayer-Bi film~\cite{Murakami2006,Liu2011,Hirahara2011} and so on. However, for practical applications, these predicted 2D materials must be placed or grown on a substrate, which would influence the electronic structure and topological properties of the hosts~\cite{Hirahara2011,Yang2012}. So we specifically consider BiCN and its allotrope $\beta-$BiCN on monolayer h-BN and MoS$_2$. (Here $\beta-$BiCN can be regarded as a bilayer Bi film passivated by -CN from oneside not both sides shown in Fig.~\ref{fig:geo-sub}(c),(d). Detailed properties are provided in part III of Supplemental Material ). As shown in Fig.~\ref{fig:geo-sub} we construct four heterostructures composed of BiCN/$\beta-$BiCN and the substrates with structural mismatches less than 5$\%$. The lattice constants of the substrates are fixed and the interlayer distances and atomic positions are relaxed to reach lowest energy structures with the vdW corrections.

\begin{figure}
  \begin{center}
  \epsfig{file=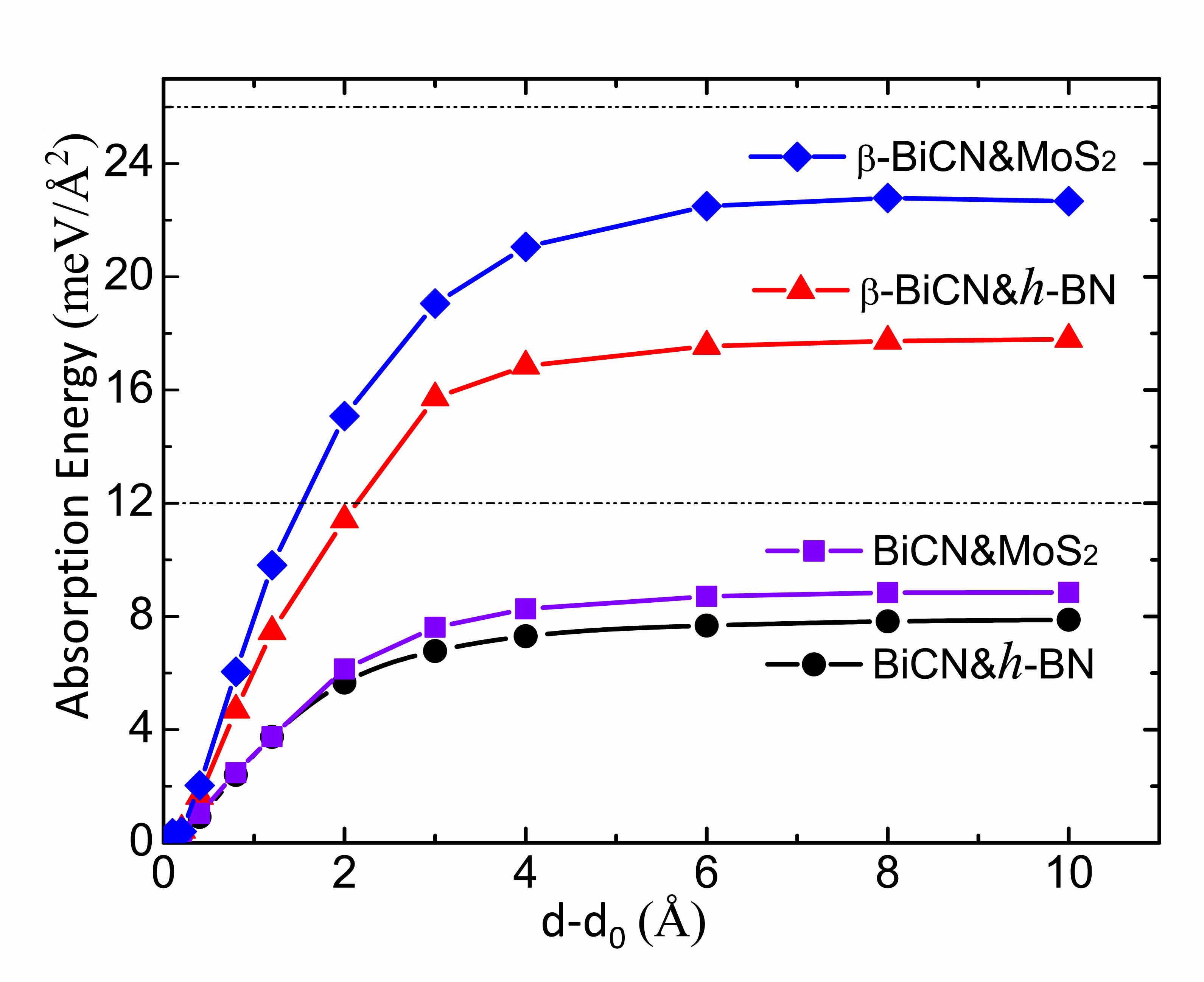,width=\textwidth}
\caption{The absorption energy curve versus the interlayer distance. d$_0$ is the interlayer distance at equilibrium state. The two black straight dash line indicate the binding energies of graphene and monolayer MoS$_2$ separated from their bulks structure respectively.\label{fig:Formation-energy}}
\end{center}
\end{figure}

To understand the interactions between BiCN/$\beta-$BiCN and the substrates, the adsorption energies, denoted as E$_{ad}$=E$_{tot}$-E$_{sub}$-E$_{host}$, are evaluated (where E$_{tot}$, E$_{sub}$, and E$_{host}$ are energies of the composite system, substrate, and  BiCN/$\beta-$BiCN  respectively). Here we calculate the dependence of the total energy E on  the interlayer distance d with respect to the  E$_{0}$ and d$_{0}$ at the equilibrium state, as plotted in Fig.~\ref{fig:Formation-energy}. When d-d$_0$ grows large enough, the E$_d$-E$_{d_0}$ will converges to a constant value, which is exactly the value of E$_{ad}$. Based on our calculation, the adsorption energies for BiCN on h-BN and MoS$_2$ are 7.78 and 8.82 meV/$\AA^2$, which are close to the vdW interlayer binding energy of graphite~\cite{Liu2012}. The adsorption energies for $\beta-$BiCN on h-BN and MoS2 are 17.78 and 22.88 meV/$\AA^2$, which are smaller than that of MoS$_2$~\cite{Bjorkman2012,Zhou2014}. Hence both can be regarded as typical vdW-type heterojuctions. Comparing the BiCN and $\beta-$BiCN on substrates, the former has relatively smaller adsorption energy. Because both sides of the Bi layer in BiCN are protected by saturated -CN, which has weak interlayer interaction with substrates. For the $\beta-$BiCN case, since the Bi layer directly faces to the substrates, it has relatively stronger interlayer interaction than the former, but still in the vdW range.

\begin{figure}
  \begin{center}
  \epsfig{file=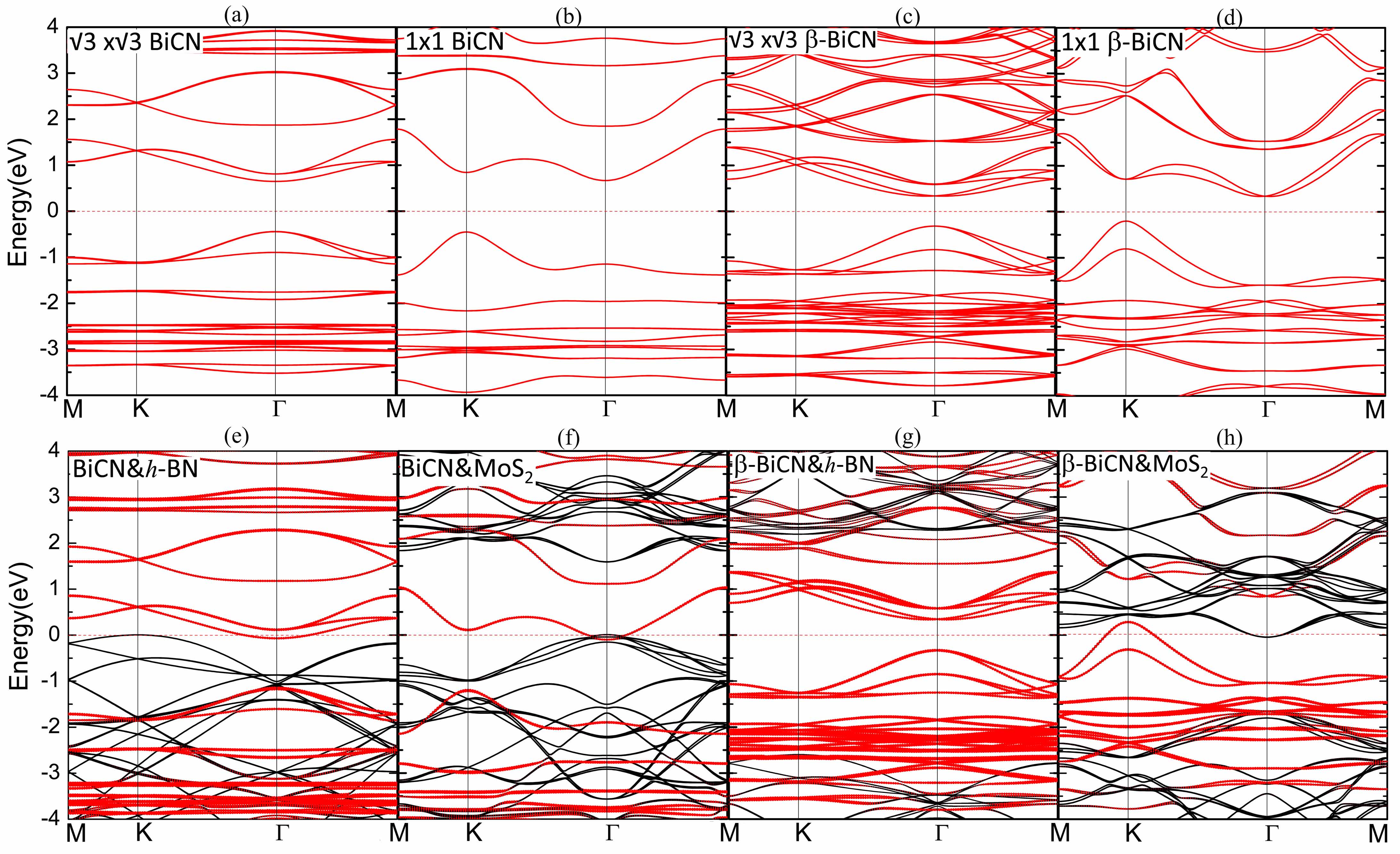,width=\textwidth}
\caption{Bandstructure for composite systems. In the bottom row (e)-(h) are the bandstructure of the four composite systems. The zero energy is set to be the Fermi level of each system. The red color and black color indicate the band contributed mainly from BiCN/$\beta-$BiCN layer and substrates respectively. In the top row (a)-(d) are the bandstructure of the pristine BiCN/$\beta-$BiCN in the same supercells as those of  (e)-(h). \label{fig:substrate-band}}
\end{center}
\end{figure}

Electronic structures for the four systems are listed in Fig.~\ref{fig:substrate-band}(e)-(h). The red dot line represents the bands from BiCN/$\beta-$BiCN and the black bands are derived from the substrates. As shown in Fig.~\ref{fig:substrate-band} (e) and (f), the bandstructures of BiCN (red) and two substrates (black) BN and MoS$_2$ are separated without being mixed. The shape of band from BiCN layer in the composite system is the same as its pristine bandstructure in the same supercell, as shown in Fig.~\ref{fig:substrate-band}(a) and (b).
They are the same as the case of $\beta-$BiCN on MoS$_2$ (Fig.~\ref{fig:substrate-band}(d) and (h)). Thus the topological properties of the BiCN on these substrates still keep invariant. It should be noticed that the fermi level crosses the valence band and conduction band from different parts of the heterojuctions and the whole systems become metallic. However we'll show that the metallic property is mainly caused by the large work function difference of BiCN/$\beta-$BiCN and the substrates, and the insulating property will recover when implying a proper gate voltage. The detailed process is discussed in part IV of Supplemental Material. The bandsturcture of the $\beta-$BiCN layer are well-located and almost intact at middle of the large band gap of substrate h-BN, and as shown in Fig~\ref{fig:substrate-band} (g) and (c),  which indicate it is an ideal topological insulating vdW heterostructure.

\section*{Discussion}
Based on DFT calculation, we predict that BiCN, i.e., bilayer Bi film passivated with -CN group, is a novel stable 2D Bi-based material, and demonstrate that it is also a new kind of 2D TI with giant SOC gap ($\sim 1$ eV) by direct calculation of the topological invariant and clear plot of  the helical edge states. Moreover, the low-energy effective Hamiltonian is given, and two candidate substrates are proposed. DFT calculation shows that the two substrates can stabilize and weakly interact with BiCN via vdW interaction and thus do not affect the electronic properties, especially the band topology of BiCN monolayer.
Bilayer Bi film with a buckling honeycomb lattice has been manufactured through molecular-beam epitaxy~\cite{Hirahara2011,Yang2012,Sabater2013}, and the  topological edge states are observed as well~\cite{Drozdov2014}. On the other hand, chemical functionalization of such 2D materials is a powerful tool to create new materials with desirable features. Although many cyanides are toxic, there are some environment friendly cyanation reagents, for example, application of a new catalytic system for cyanation reaction of various aryl halides using K$_4$[Fe(CN)$_6$] as cyanating source has been feasible experimentally~\cite{Hajipour2011}, which indicates that  K$_4$[Fe(CN)$_6$]  could also provide CN- group for the manufacture of BiCN. Therefore, it is very promising that BiCN may be synthesized by chemical reaction in solvents or by exposure of bilayer Bi film to atomic or molecular gases. 2D TIs have been proved to be strong against disorder as long as $u(1)$ gauge symmetry, and $T$  symmetries are respected on average. Breaking any symmetry may induce an exotic phenomenon. A  spontaneous magnetization can break $T$ symmetry and yield two valley-polarized quantum anomalous Hall phases with a Chern number from -1 to 1, tunable by an external magnetic field orientation~\cite{Liu2015}.  A proximity coupling to a superconductor can break $u(1)$ gauge symmetry producing a $Z_2$ topological superconductor with a Majorana Kramers pair~\cite{Fu2008,ZhangF2013}.

\section*{Methods}
Our first principle calculations are carried out by the VASP (Vienna ab-initio simulation package)~\cite{Kresse1996} within the generalized-gradient approximation (GGA) of Perdew, Burke, and Ernzerhof (PBE)~\cite{Perdew1996}. The cutoff energy of plane wave basis is set as 500 eV for the calculations. The k-meshes of $15\times15\times1$ and $1\times15\times1$ are set for 2D unit cell and 1D nanoribbon calculations, respectively.  The structures are relaxed with remaining force on each atoms less than 0.01eV/$\AA$ and the self-consistent calculations are converged  with energy difference less than 1 meV/atom for two successive steps.
A vacuum layer of 18 $\AA$  was included in Z direction to avoid the interaction of its periodic image. The phonon spectrum is calculated by QUANTUM ESPRESSO~\cite{pwscf}. The biaxial strain is implied by the change of lattice constant. The external vertical electric field is simulated by an artificially-implemented sawtooth potential in VASP. For considering substrate effects, the vdW interaction is included during the calculation by optB88-vdW correction~\cite{optB88}.

\bibliographystyle{apsrev4-1}

\begin{thebibliography}{100}

\bibitem{Hasan2010} Hasan, M. Z. \& Kane, C. L. Colloquium: Topological insulators. {\it Rev. Mod. Phys.} {\bf 82}, 3045 (2010).

\bibitem{Qi2011} Qi, X.-L. \& Zhang, S.-C. Topological insulators and superconductors. {\it Rev. Mod. Phys.} {\bf 83}, 1057 (2011).

\bibitem{kane2005a} Kane, C. L. \& Mele, E. J. Quantum Spin Hall Effect in Graphene. {\it Phys. Rev. Lett.} {\bf 95}, 226801 (2005).

\bibitem{kane2005b} Kane, C. L. \& Mele, E. J. Z$_2$ Topological Order and the Quantum Spin Hall Effect. {\it Phys. Rev. Lett.} {\bf 95}, 146802 (2005).

\bibitem{yao2007} Yao,Y. G., Ye, F., Qi, X. L., Zhang, S. C. \& Fang, Z. Spin-orbit gap of graphene: First-principles calculations. {\it Phys. Rev. B} {\bf 75}, 041401(R) (2007).

\bibitem{Min2006} Min, H. {\it et al}. Intrinsic and Rashba spin-orbit interactions in graphene sheets. {\it Phys. Rev. B} {\bf 74}, 165310 (2006).

\bibitem{Bernevig2006} Bernevig, B. A., Hughes, T. L. \& Zhang, S. C. Quantum spin Hall effect and topological phase transition in HgTe quantum wells. {\it Science} {\bf 314}, 1757 (2006).

\bibitem{Science.318.766} K$\ddot{o}$nig, M. {\it et al}. Quantum spin Hall insulator state in HgTe quantum wells. {\it Science} {\bf 318}, 766 (2007).

\bibitem{Liu2008} Liu, C. X., Hughes, T. L., Qi, X. L., Wang, K. \& Zhang, S. C. Quantum Spin Hall Effect in Inverted Type-II Semiconductors. {\it Phys. Rev. Lett.} {\bf 100}, 236601 (2008).


\bibitem{PRL.Knez.2012} Knez, I., Du, R. \& Sullivan, G. Andreev Reflection of Helical Edge Modes in InAs/GaSb Quantum Spin Hall Insulator. {\it Phys. Rev. Lett.} 109, 186603 (2012).

\bibitem{Murakami2006} Murakami, S. Quantum Spin Hall Effect and Enhanced Magnetic Response by Spin-Orbit Coupling. {\it Phys. Rev. Lett.} {\bf 97}, 236805 (2006).

\bibitem{liu_quantum_2011} Liu, C.-C., Feng, W. X. \& Yao, Y. G. Quantum Spin Hall Effect in Silicene and Two-Dimensional Germanium. {\it Phys. Rev. Lett.} {\bf 107}, 076802 (2011).

\bibitem{liu_low-energy_2011} Liu, C.-C., Jiang, H. \& Yao, Y. G. Low-energy effective Hamiltonian involving spin-orbit coupling in silicene and two-dimensional germanium and tin. {\it Phys. Rev. B} {\bf 84}, 195430 (2011).

\bibitem{Liu2011} Liu, Z. {\it et al}. Stable Nontrivial Z2 Topology in Ultrathin Bi (111) Films: A First-Principles Study. {\it Phys. Rev. Lett.} {\bf 107}, 136805 (2011).

\bibitem{Hirahara2011} Hirahara, T. {\it et al}. Interfacing 2D and 3D Topological Insulators: Bi(111) Bilayer on Bi2Te3. {\it Phys. Rev. Lett.} {\bf 107}, 166801 (2011).

\bibitem{Xu2013} Xu, Y. {\it et al}. Large-Gap Quantum Spin Hall Insulators in Tin Films. {\it Phys. Rev. Lett.} {\bf 111}, 136804 (2013).

\bibitem{Yang2012} Yang, F. {\it et al}.  Spatial and Energy Distribution of Topological Edge States in Single Bi(111) Bilayer. {\it Phys. Rev. Lett.} {\bf 109}, 016801 (2012).

\bibitem{Song2014} Song, Z. {\it et al}. Quantum spin Hall and quantum valley Hall insulators of BiX/SbX (X = H, F, Cl, and Br) monolayer with a record bulk band gap. {\it NPG Asia Materials} {\bf 6}, e147 (2014).

\bibitem{Liu2014} Liu, C.-C. {\it et al}.  Low-energy effective Hamiltonian for giant-gap quantum spin Hall insulators in honeycomb X-hydride/halide (X=N-Bi) monolayers. {\it Phys. Rev. B} {\bf 90}, 085431 (2014).

\bibitem{Zhou2014} Zhou, J.-J., Feng, W., Liu, C.-C., Guan, S. \& Yao, Y. G. Large-gap quantum spin Hall insulator in single layer bismuth monobromide Bi4Br4. {\it Nano Lett.} {\bf 14}, 4767 (2014).

\bibitem{Wang2014} Wang, Z. F., Chen, L. \& Liu, F. Tuning Topological Edge States of Bi(111) Bilayer Film by Edge Adsorption. {\it Nano Lett.} {\bf 14}, 2879 (2014).

\bibitem{Ma2015} Ma, Y., Dai, Y., Kou, L., Frauenheim, T. \& Heine, T. Robust Two-Dimensional Topological Insulators in Methyl-Functionalized Bismuth, Antimony, and Lead Bilayer Films. {\it Nano Lett.} {\bf 15}, 1083 (2015).

\bibitem{Zhang2012} Zhang, H., Freimuth, F., Bihlmayer, G., Bl$\ddot{u}$gel, S. \& Mokrousov, Y. Topological phases of Bi(111) bilayer in an external exchange field. {\it Phys. Rev. B} {\bf 86},  035104 (2012).

\bibitem{Qian2014} Qian, X., Liu, J., Fu, L. \& Li, J., Quantum spin Hall effect in two-dimensional transition metal dichalcogenides.. {\it Science} {\bf 346}, 1344-1347 (2014).

\bibitem{Huang2013} Huang, Z.-Q. {\it et al}. Nontrivial topological electronic structures in a single Bi(111) bilayer on different substrates: A first-principles study. {\it Phys. Rev. B} {\bf 88}, 165301(2013).

\bibitem{Weng2014} Weng, H., Dai, X. \& Fang, Z. Transition-Metal Pentatelluride ZrTe5 and HfTe5: A Paradigm for Large-Gap Quantum Spin Hall Insulators. {\it Phys. Rev. X} {\bf 4}, 011002 (2014).

\bibitem{Zhou2015} Zhou, L. {\it et al}. New Family of Quantum Spin Hall Insulators in Two-dimensional Transition-Metal Halide with Large Nontrivial Band Gaps. {\it Nano Lett.} {\bf 15}, 7867 (2015).


\bibitem{Wada2011} Wada, M., Murakami, S., Freimuth, F. \& Bihlmayer, G. Localized edge states in two-dimensional topological insulators: Ultrathin Bi films. {\it Phys. Rev. B} {\bf 83}, 121310(R) (2011).

\bibitem{Koroteev2008} Koroteev, Y. M., Bihlmayer, G., Chulkov, E. V. \& Bl$\ddot{\rm u}$gel, S. First-principles investigation of structural and electronic properties of ultrathin Bi films. {\it Phys. Rev. B} {\bf 77}, 045428 (2008).



\bibitem{Fu2007} Fu L. \& Kane, C. L. Topological insulators with inversion symmetry. {\it Phys. Rev. B} {\bf 76}, 045302 (2007).

\bibitem{Liu2012} Liu, Z. \emph{et. al.} , Interlayer binding energy of graphite: A mesoscopic determination from deformation. {\it Phys. Rev. B} {\bf 85}, 205418  (2012).

\bibitem{Bjorkman2012} Bj$\ddot{\rm o}$rkman, T., Gulans, A., Krasheninnikov, A. V. \& Nieminen, R. M. van der Waals Bonding in Layered Compounds from Advanced Density-Functional First-Principles Calculations. {\it Phys. Rev. Lett.}  {\bf 108}, 235502 (2012).


\bibitem{Zhou2008} Zhou, B., Lu, H.-Z., Chu, R.-L., Shen, S.-Q.\& Niu, Q. Finite Size Effects on Helical Edge States in a Quantum Spin-Hall System. {\it Phy. Rev. Lett.} {\bf 101}, 246807 (2008).


\bibitem{Sabater2013} Sabater, C. {\it et al}. Topologically Protected Quantum Transport in Locally Exfoliated Bismuth at Room Temperature. {\it Phys. Rev. Lett.} {\bf 110}, 176802 (2013).

\bibitem{Drozdov2014} Drozdov, I. K. {\it et al}.  One-dimensional topological edge states of bismuth bilayers. {\it Nat. Phys.} {\bf 10}, 664 (2014).


\bibitem{Hajipour2011} Hajipour, A. R., Karami, K., Tavakol, G. \& Pirisedigh, A. An efficient palladium catalytic system for microwave assisted cyanation of aryl halides. {\it Journal of Organometallic Chemistry} {\bf 696}, 819 (2011).

\bibitem{Liu2015} Liu, C.-C., Zhou, J.-J. \& Yao, Y. G. Valley-polarized quantum anomalous Hall phases and tunable topological phase transitions in half-hydrogenated Bi honeycomb monolayers. {\it Phys. Rev. B} {\bf 91}, 165430 (2015).

\bibitem{Fu2008} Fu L. \& Kane, C. L. Superconducting Proximity Effect and Majorana Fermions at the Surface of a Topological Insulator. {\it Phys. Rev. Lett.} {\bf 100}, 096407 (2008).

\bibitem{ZhangF2013} Zhang, F., Kane, C. L. \& Mele, E. J. Time-Reversal-Invariant Topological Superconductivity and Majorana Kramers Pairs. {\it Phys. Rev. Lett.} {\bf 111}, 056402 (2013).



\bibitem{Kresse1996} Kresse, G. \& Furthm$\ddot{\rm u}$ller, J. Efficient iterative schemes for ab initio total-energy calculations using a plane-wave basis set. {\it Phys. Rev. B} {\bf 54}, 11169 (1996).

\bibitem{Perdew1996} Perdew, J. P. Burke K. \& Ernzerhof, M. Generalized Gradient Approximation Made Simple. {\it Phys. Rev. Lett.} {\bf 77}, 3865 (1996).

\bibitem{pwscf} Giannozzi, P. {\it et al}. Quantum ESPRESSO: a modular and open-source software project for quantum simulations of materials. {\it J. Phys. Condens. Matter.} {\bf 21},  395502 (2009).

\bibitem{optB88} Klimes, J., Bowler, D. R. \& Michaelides, A., Van der Waals density functionals applied to solids. {\it Phys. Rev. B.} {\bf 83}, 195131 (2011).


\end{thebibliography}

\begin{addendum}
\item [Acknowledgements]
This work was supported by the National Natural Science Foundation of China (Grant No. 11404022), the MOST Project of China (No. 2013CB921903), Basic Research Funds of Beijing Institute of Technology (Grant No. 20141842001), and Excellent young scholars Research Fund of Beijing Institute of Technology.

\item [Author Contributions]
B.F. performed the DFT calculations. B.F., C.-C.L., Y.G., W.G. and W.S. analyzed the data and wrote the paper. C.-C.L. designed and coordinated the project.

\item [Competing Interests]
The authors declare no competing financial interests.

\item [Correspondence]
Correspondence should be addressed to Cheng-Cheng Liu.

\end{addendum}

\clearpage

\newpage

\newpage

\newpage

\newpage
\begin{figure}
  \begin{center}
  \epsfig{file=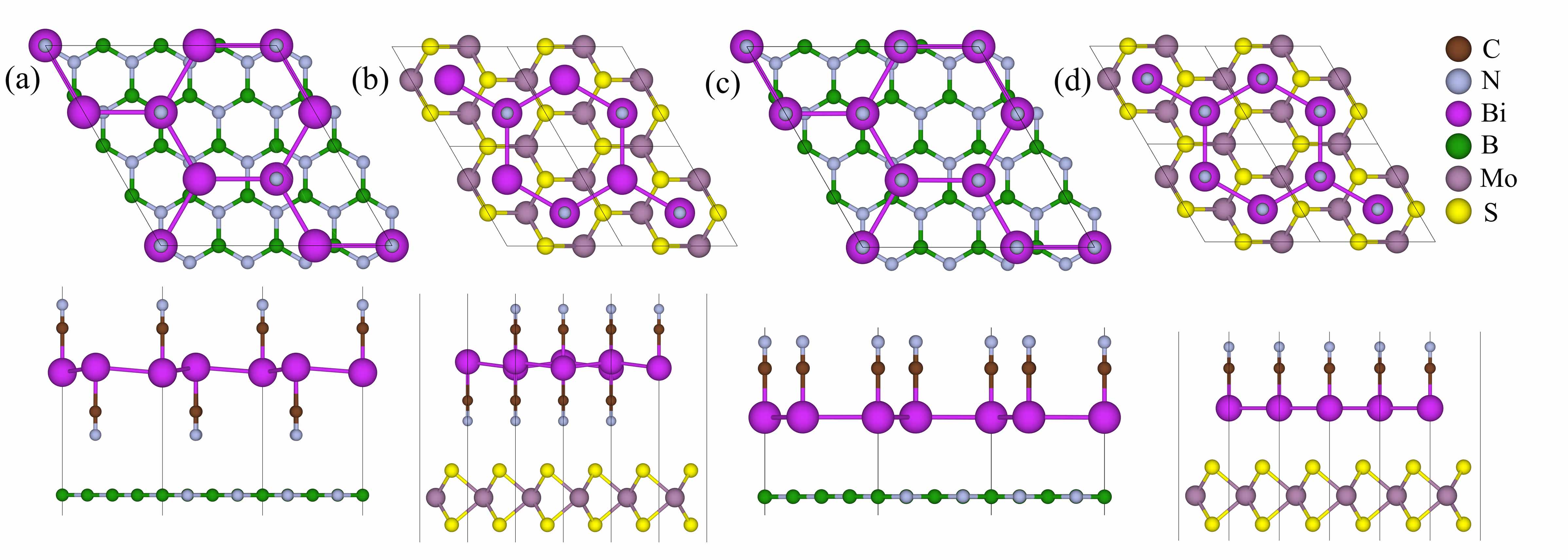,width=\textwidth}
\caption{Optimized atomic structures of four composite systems with top view (top panel) and side view (bottom panel). (a) $\sqrt{3}\times\sqrt{3}$ BiCN on $4\times4$ h-BN  with lattice mismatch of -4.5$\%$ and interlayer distance of 3.01 $\AA$ (b) $1\times1$ BiCN on $\sqrt{3}\times\sqrt{3}$ MoS$_2$ with lattice mismatch of 0.6 $\%$ and interlayer distance of 2.85 $\AA$ (c) $\sqrt{3}\times\sqrt{3}$ $\beta-$BiCN on $4\times4$ with lattice mismatch of 0.5 $\%$ and interlayer distance of 3.48 $\AA$ (d) $1\times1$ $\beta-$BiCN on $\sqrt{3}\times\sqrt{3}$ MoS$_2$ with lattice mismatch of -4.8 $\%$ and interlayer distance of 3.23 $\AA$.\label{fig:geo-sub}}
\end{center}
\end{figure}

\newpage

\newpage

\end{document}